\documentclass[preprint]{aastex}

\usepackage{color}

\shorttitle{Outbursting Comet P/2010 V1 (Ikeya--Murakami)}
\shortauthors{Ishiguro et al.}

\begin{document}

\title{Outbursting Comet P/2010 V1 (Ikeya--Murakami):
A Miniature Comet Holmes}

\author{Masateru \textsc{Ishiguro}\altaffilmark{1}}
\affil{Department of Physics and Astronomy, Seoul National University,\\
Gwanak, Seoul 151-742, Republic of Korea}

\author{David Jewitt}
\affil{Department of Earth, Planetary and Space Sciences,
University of California at Los Angeles, \\
595 Charles Young Drive East, 
Los Angeles, CA 90095-1567\\
Department of Physics and Astronomy,
University of California at Los Angeles, \\
430 Portola Plaza, Box 951547,
Los Angeles, CA 90095-1547}

\author{Hidekazu \textsc{Hanayama}}
\affil{Ishigakijima Astronomical Observatory, National Astronomical
Observatory of Japan,\\ Ishigaki, Okinawa 907-0024, Japan}

\author{Fumihiko \textsc{Usui}}
\affil{Department of Astronomy, Graduate School of Science, 
The University of Tokyo,\\ 7-3-1 Hongo, Bunkyo-ku, Tokyo 113-0033, Japan}

\author{Tomohiko \textsc{Sekiguchi}}
\affil{Asahikawa Campus, Hokkaido University of Education, 9 Hokumon,
Asahikawa 070-8621, Japan}

\author{Kenshi \textsc{Yanagisawa} and Daisuke \textsc{Kuroda}}
\affil{Okayama Astrophysical Observatory, National Astronomical
Observatory of Japan, Asaguchi, Okayama 719-0232, Japan}

\author{Michitoshi \textsc{Yoshida}}
\affil{Hiroshima Astrophysical Science Center, Hiroshima University,
1-3-1 Kagamiyama, Higashi-Hiroshima, Hiroshima 739-8526, Japan} 

\author{Kouji \textsc{Ohta}}
\affil{Department of Astronomy, Kyoto University, Kyoto 606-8502, Japan}

\author{Nobuyuki \textsc{Kawai}}
\affil{Department of Physics, Tokyo Institute of Technology
2-12-1 Ookayama, Meguro-ku, Tokyo 152-8551, Japan}

\author{Takeshi \textsc{Miyaji}}
\affil{Ishigakijima Astronomical Observatory, National Astronomical
Observatory of Japan,\\ Ishigaki, Okinawa 907-0024, Japan}

\author{Hideo \textsc{Fukushima}\altaffilmark{2}, and
Jun-ichi \textsc{Watanabe}\altaffilmark{2}}
\affil{National Astronomical Observatory of Japan, Mitaka, Tokyo,
181-8588, Japan}

\altaffiltext{1}{Visiting Scientist, Department of Earth, Planetary
and Space Sciences, University of California at Los Angeles, 595 Charles Young
Drive East, Los Angeles, CA 90095-1567, USA}

\altaffiltext{2}{Ishigakijima Astronomical Observatory, National Astronomical
Observatory of Japan, Ishigaki, Okinawa, 907-0024, Japan}

\begin{abstract}
Short-period comet P/2010 V1 (Ikeya-Murakami, hereafter ``V1'') was
discovered visually by two amateur astronomers. The appearance of the comet was peculiar, consisting of
an envelope, a spherical coma near the nucleus and a tail 
 extending in the anti-solar direction. We investigated the brightness
 and the morphological development of the comet by taking optical images
 with ground-based telescopes. Our observations show that  V1 experienced a large-scale explosion between UT 2010
October 31 and November 3.  The color of the comet was
consistent with the Sun ($g'-R_\mathrm{C}$=0.61$\pm$0.20,
$R_\mathrm{C}-I_\mathrm{C}$=0.20$\pm$0.20, and
$B-R_\mathrm{C}$=0.93$\pm$0.25), suggesting that dust particles were
responsible for the brightening. We used a dynamical model
 to understand the peculiar morphology, and found that 
 the  envelope consisted of small grains (0.3--1 \micron)
expanding at a  maximum speed of 500$\pm$40 m s$^{-1}$,  while the tail
and coma were composed of a wider range of dust particle sizes (0.4--570\micron) and expansion speeds 7--390 m s$^{-1}$.  
The total  mass of ejecta is $\sim$5$\times$10$^{8}$ kg and  kinetic energy $\sim$5$\times$10$^{12}$ J. These values are much smaller than in the historic outburst of 
17P/Holmes in 2007, but the energy per unit mass (1$\times$10$^4$ J
kg$^{-1}$) is comparable. The energy per unit
mass is about 10\% of the energy released during the crystallization of
amorphous water ice suggesting that crystallization of buried amorphous
ice can supply the mass and energy of the outburst ejecta.  
\end{abstract}

\keywords{interplanetary medium --- comets --- comets:
individual (P/2010 V1) --- solar system}

\section{INTRODUCTION}
\label{sec:introduction}

\begin{figure}
 \epsscale{0.8}
   \plotone{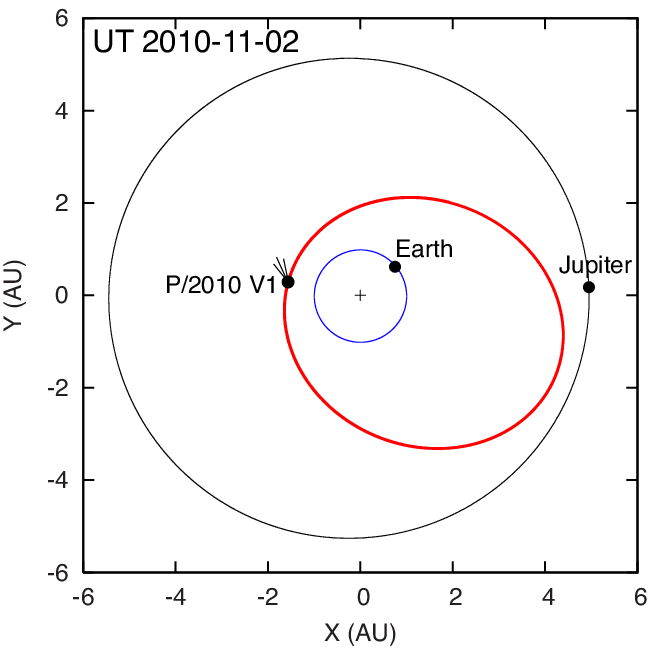}
  \caption{Orbits of P/2010 V1 (Ikeya--Murakami) projected on the ecliptic plane.
 Large ellipse is the orbit of the Jupiter, and smallest ellipse in the
 orbit of the  Earth, respectively. Cross denotes the position of the
 Sun, and filled circles mean the positions of the comet, Earth, and
 Jupiter on UT 2010 November 2, a potential date of the
 outburst. This is shown in the heliocentric ecliptic coordinate, that
 is, the x-axis points from the Sun toward the vernal equinox and
 y-axis completes the right-hand ecliptic coordinate system.}
 \label{fig:orbit}
\end{figure}

Periodic comet, P/2010 V1 (Ikeya-Murakami, hereafter V1) was
independently discovered by two amateur astronomers in Japan,
Mr. Kaoru Ikeya and Dr. Shigeki Murakami, in early 2010 November
\citep{Nakano2010a}. They reported the comet to be at magnitude 8--9 at the time of
discovery. Later, the orbital elements (semimajor
axis $a$=3.083 AU, eccentricity $e$=0.488, and inclination
$i$=9.38\arcdeg) showed that V1 is a short period comet with an orbital period of 5.41 years \citep{Williams2010}. Figure
\ref{fig:orbit} shows the orbit projected on the ecliptic plane.
It has a Tisserand parameter with respect to 
Jupiter, T$_J$=3.013, slightly larger than 3. Such comets are sometimes classified
as Encke-type comets (2P/Encke has $T_J$ = 3.026) rather than Jupiter-family comets, for which 2 $\le T_J<$ 3
\citep{Levison1997}. 
Despite its short orbital period and considerable brightness at the time of  discovery, it is
interesting to note that V1 had not been previously detected.

To date, there are no published reports to characterize the physical properties of V1.
Images taken by amateur astronomers showed interesting features. The
comet was enveloped by a spherical cloud and the overall appearance was
reminiscent of historic cometary outbursts in 17P/Holmes. To characterize
the physical properties, we obtained monitoring observations and compared them with a model based on the dynamics of dust grains.


\section{OBSERVATIONS AND DATA REDUCTION}
\label{sec:observation}

The data presented in this study were obtained with three
telescopes: the Ishigakijima Astronomical Observatory Murikabushi
1.05-m telescope (hereafter IAO), the Keck I 10-m telescope (Keck-I),
and the Indian Institute of Astrophysics 2.0-m Himalayan Chandra
telescope (HCT). A journal of the observations is given in Table
1. Details of the data acquisition and reduction are given in the
following.

Long-term monitoring observations of V1 were taken at
IAO, in Okinawa, Japan with the Murikabushi 1.05-m 
Ritchey-Chr\'etien telescope (F/12) with a focal reducer and MITSuME, a
system to take contemporaneous images with three different filters of
SDSS $g'$, Johnson--Cousins $R_\mathrm{C}$, and
$I_\mathrm{C}$-band. Each of the three cameras utilizes an Alta U6 (Apogee Instruments Inc.) CCD with  array size of 1024 $\times$ 1024 pixels and
with pixel size of 24 $\times$ 24 \micron.
The effective wavelengths and the full width at half-maximum (FWHM) are
 $\lambda_e$=4830\AA~and  $\Delta \lambda$=1340\AA~($g'$), 
 $\lambda_e$=6550\AA~and  $\Delta \lambda$=1210\AA~($R_\mathrm{C}$),  and
 $\lambda_e$=7990\AA~and  $\Delta \lambda$=1570\AA~($I_\mathrm{C}$).
In this configuration, the pixel size projected on the sky was 0.72\arcsec~and the field of view was
12.3\arcmin$\times$12.3\arcmin. The observations were made using
non-sidereal tracking in sky conditions that were variable through our
observation runs.

Multiband snapshots were obtained with the 10-m Keck I telescope
atop Mauna Kea on UT 2011 January 30. Images were taken using the Low
Resolution  Imaging Spectrometer (LRIS) camera \citep{Oke1995},
which houses red and blue optimized CCDs separated by a dichroic filter
(we used the 460 dichroic, which has 50\% transmission 
at 4875 \AA~).  The 
image scale on both cameras was 0.135\arcsec~per pixel and the available
field-of-view was 5.3\arcmin$\times$7.3\arcmin. The telescope was tracked at sidereal rates owing to temporary failure of the Keck guider control software.   We secured two sets of images simultaneously in the B-band
($\lambda_e$=4370 \AA~and $\Delta \lambda$=900\AA~)
and R-band ($\lambda_e$=6800 \AA~and $\Delta \lambda$=1270\AA~) filters, with exposures of
25 s and 20 s for the first set and 250 s and 200 s for the second set, respectively. We used the first set
because the comet was trailed due to the sidereal tracking in the
second set.  The sky above Mauna Kea
was photometric.

The last observation for V1 was carried out on UT 2011 March 29 with the 
2.0-m Ritchey-Chr\'etien HCT located at 4500 meters in the
Himalayan region, India. It is operated by the Indian Astronomical Observatory,
the Indian Institute of Astrophysics, (IIA). We employed the Himalaya Faint Object
Spectrograph (HFOSC) $2048\times 4096$ pixel CCD camera with
$R_\mathrm{C}$-band filter  ($\lambda_e$=6550 \AA~and $\Delta \lambda$=1450\AA) 
at the f/9 Cassegrain focus of the telescope. The image scale on the camera was
0.296\arcsec~per pixel and the available field-of-view was $10\arcmin\times 10\arcmin$. 
The observation was conducted in a crowded region of stars at the
galactic longitude and latitude of 354.4\arcdeg~ and -1.5\arcdeg. We
could not detect the comet with the HCT but used these data to place an
upper-limit to the brightness. 

The  raw images were reduced in the standard manner for CCD data.
The bias data were obtained at intervals throughout each night.
We used median-stacked data frames to construct flat-field images with
which to correct for pixel-to-pixel variation in CCD response and vignetting.
Flux calibration was obtained using standard stars in the Landolt catalog when 
available \citep{Landolt1992,Landolt2009}, otherwise we used field stars 
listed in the USNO--B1.0 catalog \citep{Monet2003}. We employed
WCSTools to transform CCD pixel coordinates into celestial coordinates
\citep{Mink1997}. The estimated astrometric accuracy was about
0.4\arcsec, which is good enough to argue the position angle and
morphology of dust structure in the following section. To remove cosmic rays and background
objects such as galaxies and stars in IAO and HCT data, we followed a technique
described in \citet{Ishiguro2007} and  \citet{Ishiguro2008}. 
The technique is useful only when a number of exposures were acquired.
For the Keck-I image, we did not delete stars because   only one set
of exposures was available.


\section{RESULTS}

\subsection{THE COLOR}
\begin{figure}
 \epsscale{0.8}
   \plotone{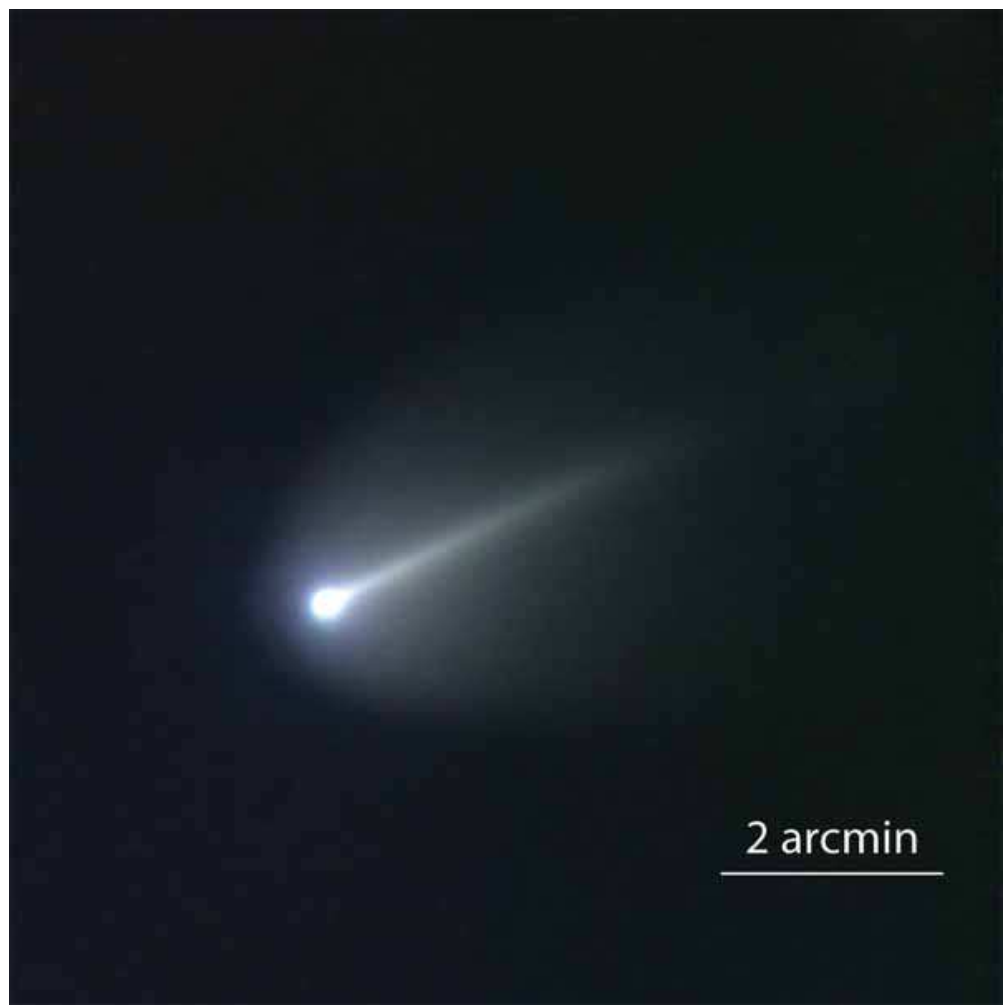}
  \caption{$g'R_\mathrm{C}I_\mathrm{C}$-band composite color image of
 V1 taken on UT 2010 November 9. We allocated $g'$-band
 image as blue, $R_\mathrm{C}$-band image as green, and
 $I_\mathrm{C}$-band image  as red to make the color image. The
 Celestial North is up and  Celestial East to the left. The
 field-of-view of the image is 9\arcmin$\times$9\arcmin.}
 \label{fig:color_image}
\end{figure}

Figure \ref{fig:color_image} shows a 
false-color composite image taken on UT 2010 November 9. In the Figure, we assigned a $g'$-band image to the blue color, a $R_\mathrm{C}$-band
image to the green color, and an $I_\mathrm{C}$-band image to the red color,
respectively. At a glance, the comet has a whitish color suggesting that the intensity
distribution is similar among these three bands. We derived the apparent
magnitudes of the entire cloud on UT 2010 November 9 as $g'$=10.14$\pm$0.13,
$R_\mathrm{C}$=9.53$\pm$0.14, and $I_\mathrm{C}$=9.33$\pm$0.14. In
addition, we measured the color of near-nucleus dust within an aperture
of 1\arcsec~in radius on 2011 January 30 using Keck-I images and derived
$B-R_\mathrm{C}$=0.93$\pm$0.25. The color indices of the cloud,
$g'-R_\mathrm{C}$=0.61$\pm$0.20 and
$R_\mathrm{C}-I_\mathrm{C}$=0.20$\pm$0.20 on November 9, and 
$B-R_\mathrm{C}$=0.93$\pm$0.25 on January 30, are consistent with 
those of the Sun, that is, ($g'-R_\mathrm{C}$)$_\odot$=0.65 \citep{Kim2012},
($R_\mathrm{C}-I_\mathrm{C}$)$_\odot$=0.33, and
($B-R_\mathrm{C}$)$_\odot$=1.00 \citep{Holmberg2006}. It is,
therefore, natural to think that scattered sunlight by dust particles
accounted for a large fraction of the flux in the cloud.  

\begin{figure}
 \epsscale{0.6}
   \plotone{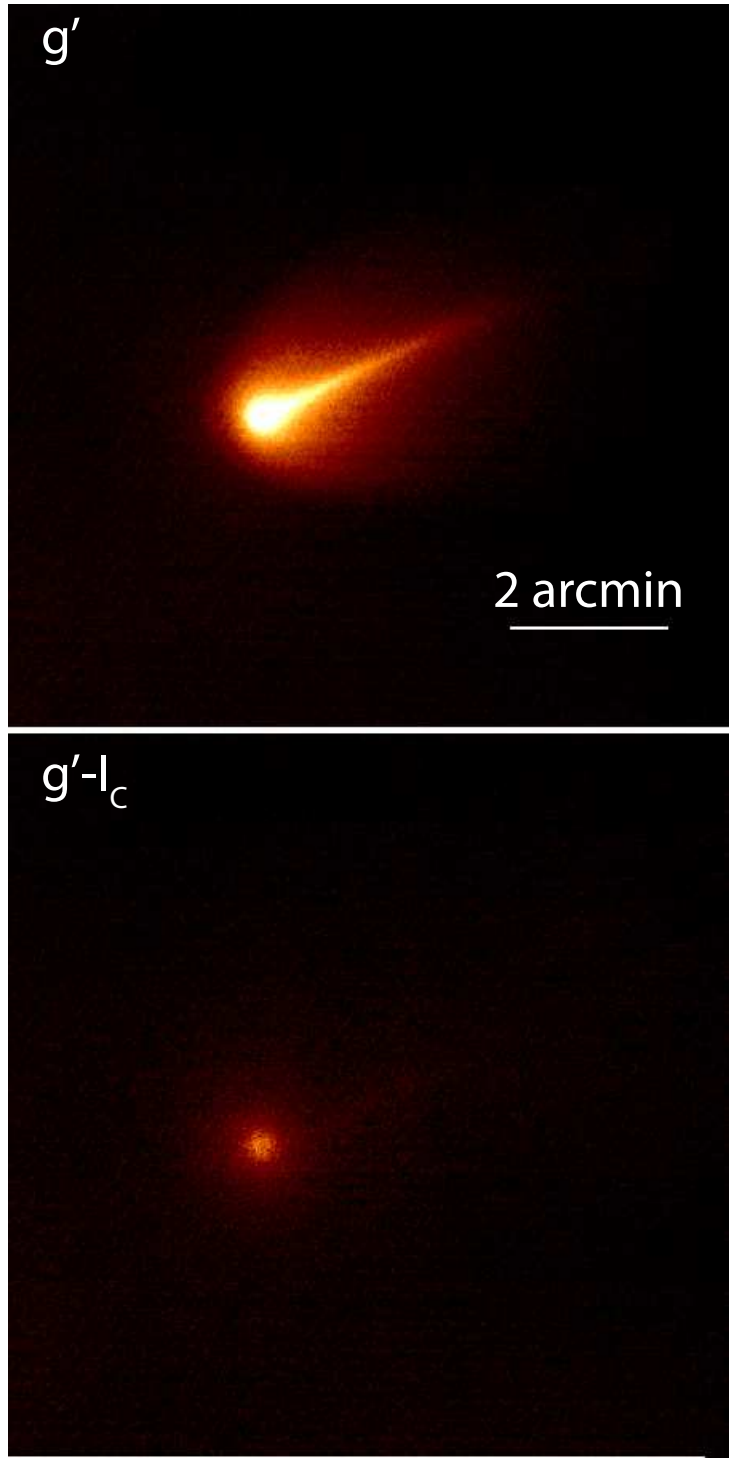}
  \caption{$g'$-band image (top) and differential intensity image between $g'$ and
 $R_\mathrm{C}$-bands image (i.e.~$g'$-band intensity minus $R_\mathrm{C}$-band intensity, bottom).
 The contribution of the spherical
 coma is about 10\% of the signal near the nucleus. This image was taken
 on UT 2010 November 9. The
 orientation of these images are the same as Figure
 \ref{fig:color_image}, that is, Celestial North is up and  Celestial 
 East to  the left. The field-of-view of the image is
 9\arcmin$\times$9\arcmin. } 
 \label{fig:subtract_image}
\end{figure}

Careful investigation  enables us to find subtle
differences between images taken in different filters. Based on inspection of the spectra of other comets, we assumed that the observed
$I_\mathrm{C}$-band intensity is wholly due  to dust continuum,
and then extracted a signal from other filters associated with emission lines from gaseous
atoms and molecules excited to fluorescence by sunlight. 
Figure \ref{fig:subtract_image} shows the differential images on UT 2010
November 9. We forced a match to the brightness level of the observed envelope
in each band in order to subtract the dust continuum. The comparison shows a
spherical cloud in the $g'$-band image, centered on the nucleus. This cloud was not clear
in $R_\mathrm{C}$-band (less than a few percent of dust continuum). Spherical structures are often detected in comets, 
where they are attributed to C$_2$ (4500--4800\AA, 4900--5200\AA, and 5300--5600\AA)
and NH$_2$ (4900--5000\AA, $\sim$5200\AA, $\sim$5400\AA, $\sim$5700\AA, and $\sim$6000\AA) 
\citep{Capria2010,Brown1996,Combi1980}. For the subsequent analysis, we used the $R_\mathrm{C}$-band 
images because they are more sensitive than $I_\mathrm{C}$-band images while remaining
less contaminated by gaseous emission than are $g'$-band images. 

\subsection{TIME-EVOLUTION OF MORPHOLOGY}

\begin{figure}
 \epsscale{1.0}
   \plotone{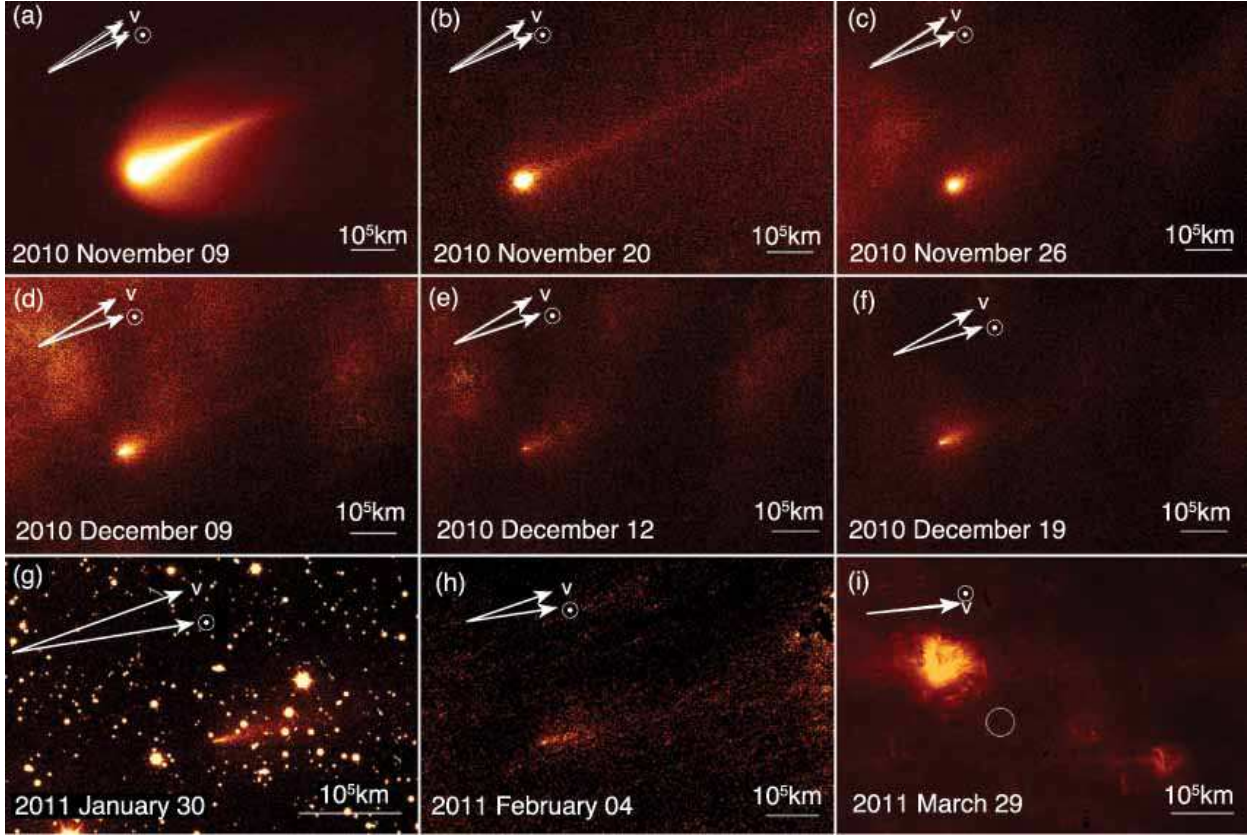}
  \caption{Time-series $R_\mathrm{C}$-band images of V1
 on  (a) UT 2010 November 9, (b) November 20, (c)  November 26, (d)
 December  9, (e) December 12, (f) December 19, (g)  2011 January 30,
 (h) February 04, and (i) March  29 in arbitrary brightness
 scale. These images  have  the standard   orientation in the sky,  that
 is, North is up and East  is to  the  left. The field of view of the
 image is 9\arcmin$\times$6\arcmin~ for (a)--(f) and (h)--(i) and
 4.5\arcmin$\times$3\arcmin~ for (i). The cardinal directions are marked,
 as are the projected anti-solar direction ($\odot$) and the projected
 negative heliocentric velocity vector (V). The predicted position of the
 comet in (h) was illustrated by a circle whose radius corresponds to
 root-sum-of-squares of the 3-standard deviation plane-of-sky error
 ellipse based on NASA/JPL ephemeris generator. There was no detectable
 object brighter than 20.0 mag in the predicted position. Note that the
 background of skies in (c)--(e) were largely contaminated by off-axis
 scattered light from Venus (smudge-like features in upper left of the
 images) while patchy features in (i) are remnants of bright stars and
 diffuse galaxies.}  
 \label{fig:time_image}     
\end{figure}

As  mentioned above, the optical image showed a unique morphology of the dust cloud
consisting of an envelope, a near-nucleus coma, and a tail
(see Figure \ref{fig:color_image}). Figure \ref{fig:time_image} shows 
the time-series $R_\mathrm{C}$-band images of V1 from UT 2010 November 9 to
UT 2011 March 29. Note that the smudge-like features in Figure
\ref{fig:time_image} (c)--(e) are artifacts of off-axis scattered
light from Venus. The envelope was clear in the first image (Figure
\ref{fig:time_image}(a)), hardly visible in the second image (Figure
\ref{fig:time_image}(b)), and undetectable after the third day of our
observation. On the other hand, the near-nucleus coma and the tail
persisted until UT 2011 February 4 (Figure
\ref{fig:time_image}(a)--(h)). Finally, nothing was detected on UT 2011
March 29 (Figure \ref{fig:time_image}(h)). We 
show the predicted position of the comet in Figure
\ref{fig:time_image}(h) using NASA/JPL's Horizons ephemeris
generator\footnote{http://ssd.jpl.nasa.gov/}. No object
brighter than 20.0 mag was detected. Assuming the
geometric albedo of 0.04 (typical of comets), we determined an upper
limit of the nuclear radius at $\approx$1850 m. 

\begin{figure}
 \epsscale{0.8}
   \plotone{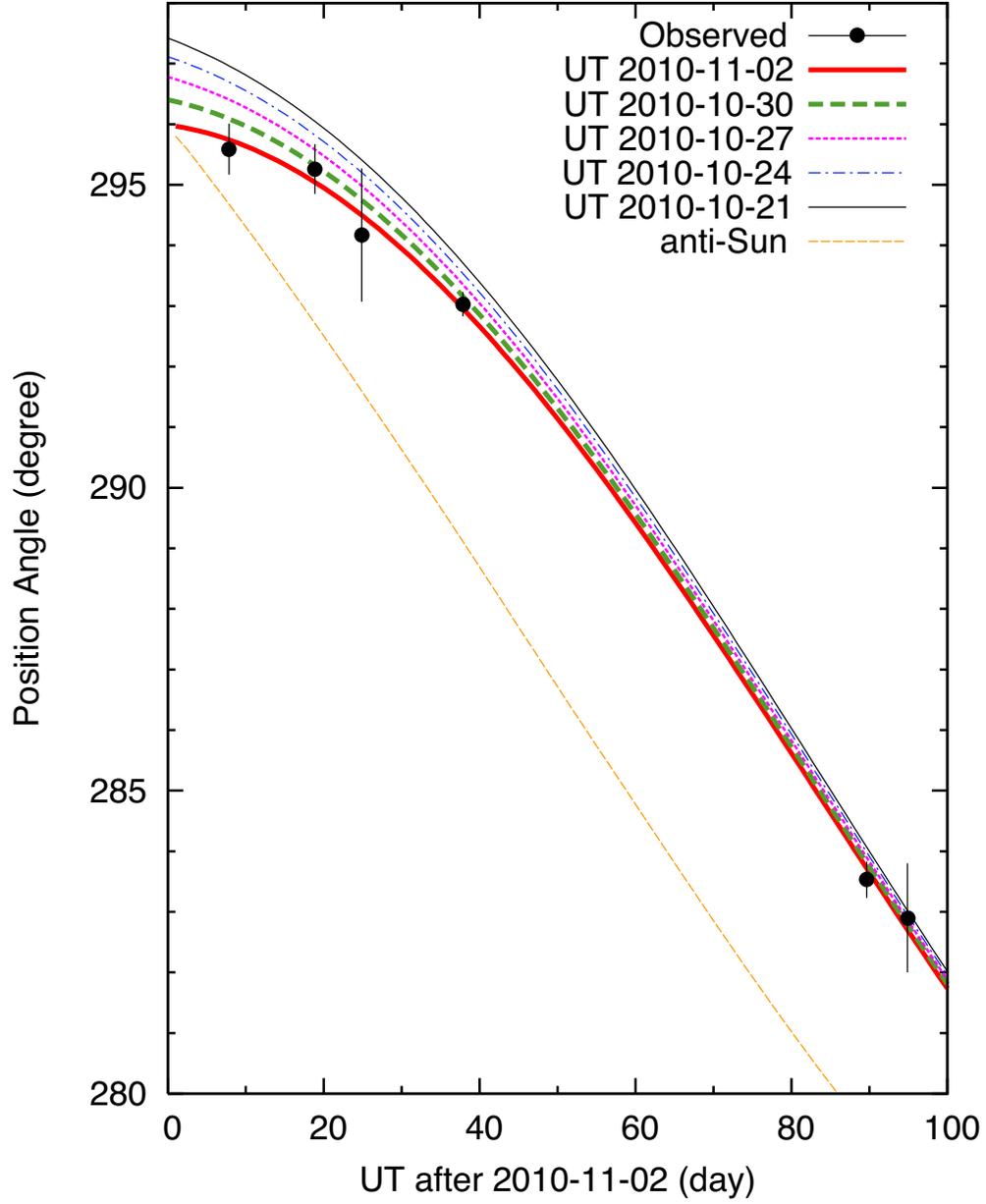}
  \caption{Position angles of the dust tail as a function of time
 showing changes caused by the viewing geometry. The measured position
 angles of the tail are indicated by filled circles with error bars
 denoting one standard deviation. Calculate position angles of different
 synchrones are also shown, labelled by the ejection time.}
 \label{fig:PA}     
\end{figure}

In Figure \ref{fig:time_image}, we see that the orientation of the tail
changed with time. To measure the position angles of the tail, we first
applied the Larson--Sekanina 
filter \citep{Sekanina1984} in order to enhance
fine-scale structures. We obtained profiles
perpendicular to the projected orbit by averaging over 15--100
pixels parallel and 1--3 pixels perpendicular to the orbit. To each profile
we fitted a Gaussian function. We then fitted a linear function to the
peak of the Gaussian versus the distance from the nucleus. The slope and
root-mean-square of the slope give us the position angle of the tail and
the corresponding error bars \citep{Jewitt2010}. We plot the position
angles as a function of the observed time (Figure \ref{fig:PA}).  We
initially compared these position angles with that of the anti-Sun vector
(the extended Sun to  comet radius vector as seen in the
plane-of-sky), but found that the observed position angles significantly
deviated from the anti-Sun vector. In addition, we compared them with
synchrones, that is, the loci of dust particles emitted at  specific
dates with zero ejection velocity. 
In Figure \ref{fig:PA}, it is clear that synchrones reproduce the  position angles over the full range of dates observed, consistent with impulsive, rather than continuous, emission of dust.  Specifically, we found best-fitting synchrone dates in the range from
UT 2010 October 31 to November 3. These dates are consistent with a
reported non-detection by Mr. Ikeya on November 1.8, one day before discovery of the comet on November 2.8
\citep{Nakano2010b}. We conclude that an outburst occurred on V1
between UT 2010 October 31 and November 2.8, and most likely between November 1.8 and 2.8. In the remainder of this paper, we adopt UT 2010 November 2 as the time of outburst, after confirming that  uncertainties in this date by up to 2 days  do not materially change the interpretation below.

\subsection{PHOTOMETRY OF THE NEAR-NUCLEUS COMA}
\label{sec:free_expansion}

The near-nucleus coma was visible as an approximately circular
dust cloud. We obtained aperture photometry to study the material close
to the nucleus with the aim of monitoring the comet's continued activity
after its explosion. The photometry was performed using the {\it APPHOT}
package in 
IRAF, which provides the magnitude within synthetic circular apertures
projected onto the sky. We used apertures of fixed physical radius at
the comet. A circular aperture of projected radius 15,000 km was
used, corresponding to angular radii 8.9\arcsec--12.3\arcsec.  The apertures were
large enough to be unaffected by seeing variations from night to night.  Table 2 lists the measured $R_\mathrm{C}$-band magnitudes, $m_R$. 

We represent the absolute magnitude (i.e.~the magnitude at a
hypothetical point at unit 
heliocentric distance and observer's  distance and at zero solar phase
angle),  by:   

\begin{eqnarray}
m_R(1,1,0)=m_R - 5~\log(r_h \Delta) - \beta \alpha,
\label{eq:eq1}
\end{eqnarray}

\noindent where $\Delta$ and $r_h$ are the observer's distance and the
heliocentric distance in AU, $\beta$ is the phase coefficient and $\alpha$ is the solar phase angle in
degree. We used $\beta=0.035$ mag deg$^{-1}$ as determined from measurements of other comets \citep{Lamy2004}.

\begin{figure}
 \epsscale{0.8}
   \plotone{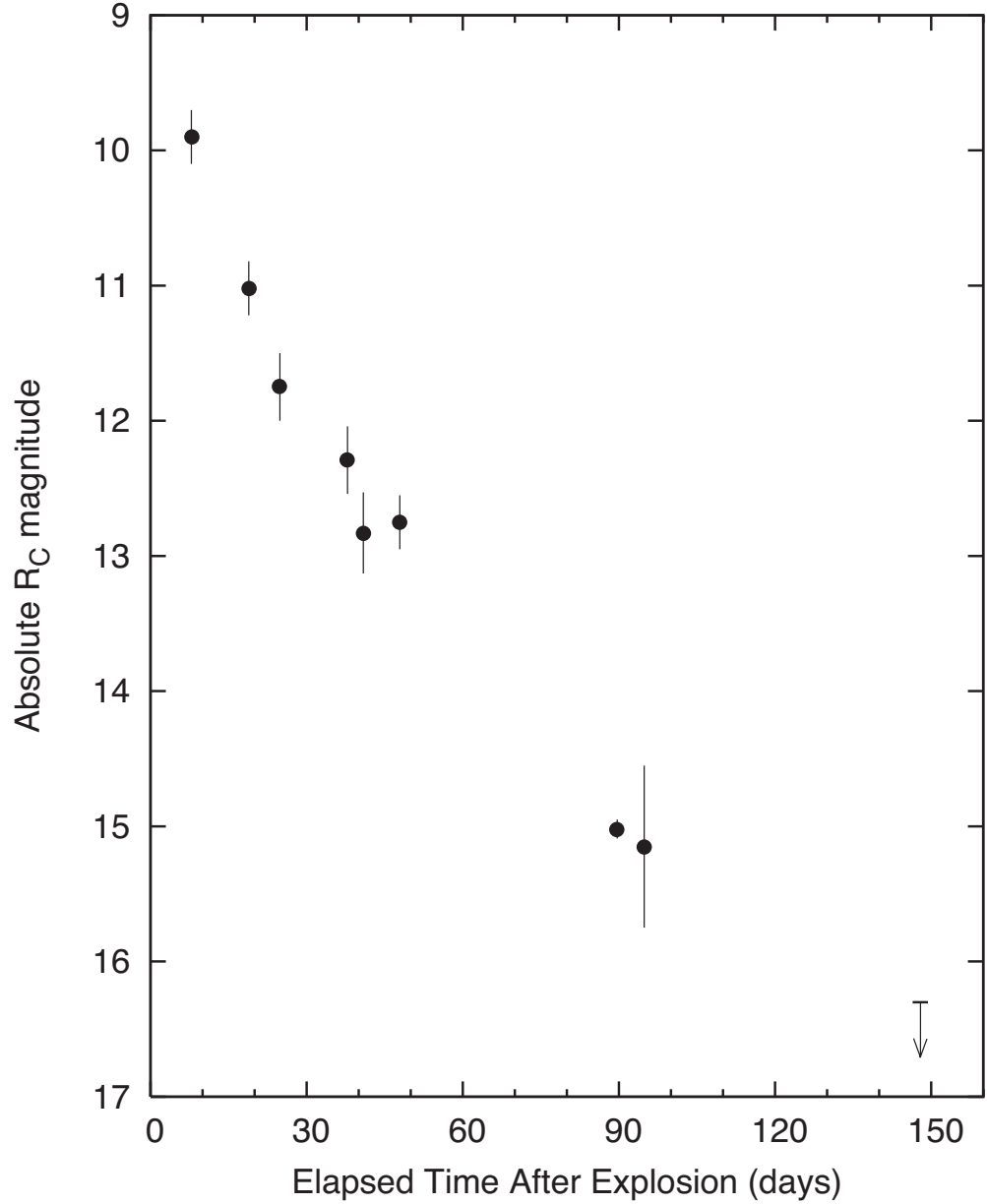}
  \caption{$R_\mathrm{C}$-band photometric evolution of the V1's inner
 coma during UT 2010 November 9 and 2011 March 29 with a  15,000 km radius
 aperture. The horizontal axis is the elapsed time in  day since the
 potential outburst date (UT 2010 November 2).  The  magnitude decreased by 
 0.06 a day over the period. Because no significant signal
 was detected on 2011 March 29, we show the upper limit of the
 magnitude.} 
 \label{fig:mag}     
\end{figure}

Figure \ref{fig:mag} shows the absolute $R_\mathrm{C}$-band magnitude of
the dust coma as a function of time after UT 2010 November 2 (i.e.~the
day of the explosion). We show an upper limit from the last data
taken with HCT. In the figure, we did not subtract the contribution to the flux from the nucleus.  
This contribution is unknown but probably negligible compared with
dust cloud. We see that the coma magnitude decreased by
$\sim$5 magnitude (a factor of $\sim$100) over $\sim$80 days.
The fading rate of V1 ($\sim$0.06 mag day$^{-1}$) is slightly slower but
approximately consistent with that of 17P/Holmes (0.08 mag day$^{-1}$
when measured through a small photometry aperture, 2500km,
\citep{Stevenson2012}).

\begin{figure}
 \epsscale{0.8}
   \plotone{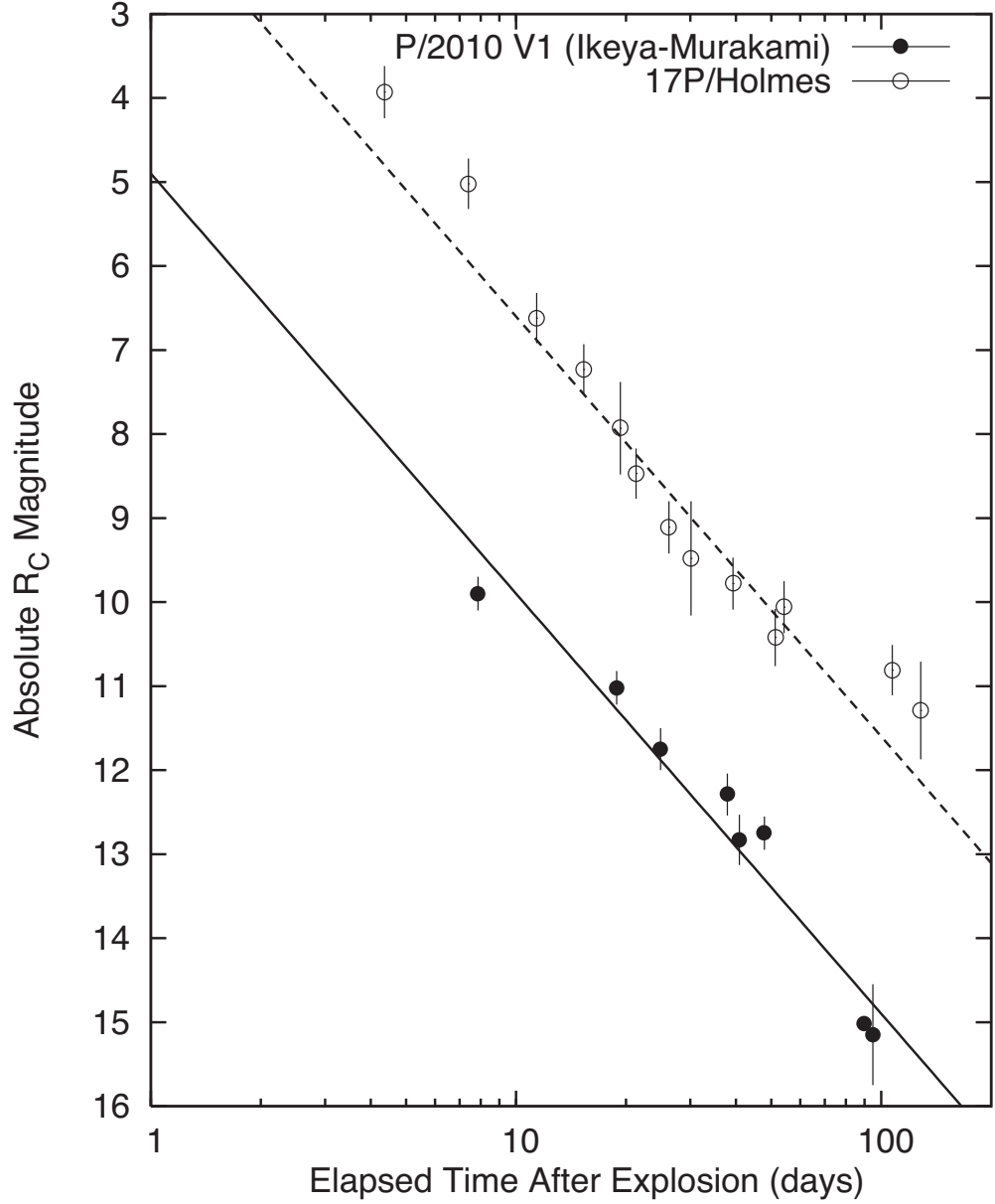}
  \caption{Comparison between $R_\mathrm{C}$-band magnitude of the V1's
 inner coma and a free expansion model (lines) in logarithm scale. For
 comparison, we show $R_\mathrm{C}$-band magnitude of 17P/Holmes with
 the same physical  radius, where we assumed the onset  time of the
 outburst on 2007 October 23.3 \citep{Hsieh2010}.}
 \label{fig:mag_comp}     
\end{figure}

To understand the magnitude profile in Figure \ref{fig:mag}, we
contrived a simple free expansion model in which dust particles expanded at
a constant speed without any acceleration. In the model, we assumed that 
dust particles reached the projected aperture radius of our photometry
(i.e.~15,000 km) throughout our observation. To validate the assumption,
dust particles should have the initial speed $>$25 m s$^{-1}$ to 
reach the projected radius on UT 2010 November 9 (we justify the
assumption of the ejection speed in the following section). The number
density of the cloud within the 15,000 km sphere decreases inversely
with the cube of elapsed time. On the other hand, 
the length along the line-of-sight increases in direct proportion to
elapsed time. As the result, the total number of particles within the
15,000 km sphere decreases as the inverse square of elapsed time. It
suggests that the magnitude of the dust coma within the fixed physical
radius can be described as $m_R(1,1,0)=5\log(\Delta t)+m_0$, where 
$\Delta t$ denotes the elapsed time and $m_0$ is a constant. We draw the
line of  $m_R(1,1,0)=5\log(\Delta t)+m_0$ in Figure \ref{fig:mag_comp}
adjusting $m_0$. 

For comparison, we plot photometric results for
17P/Holmes also obtained with a circular aperture of projected radius 15,000 km (Table \ref{tab:17Pphotometry}). 
The 17P/Holmes data were acquired at Kiso Observatory with the 2KCCD
camera attached to the 1.05-m Schmidt telescope, and obtained from the
public data archive, SMOKA.
Although it is a crude model to describe the
free expansion and there could be complicating factors such as dust 
disaggregation \citep{Li2011,Sekanina1982} and sublimation of icy grains
\citep{Stevenson2012,Yang2009} as well as acceleration by solar
radiation pressure, the fading trend is well matched by the free
expansion model. We conclude that the bulk of the dust in V1 was ejected impulsively. 

\section{Discussion}

\subsection{DUST DYNAMICAL MODEL}
For a better understanding of the unique morphology on UT 2010 November 9,
we created model images of V1 based on a dynamical theory of dust
grains. The dynamics of dust grains are determined both by the ejection
speed ($V_{ej}$) and by the ratio of radiation pressure acceleration to
solar gravity ($\beta_{rp}$). For spherical particles, $\beta_{rp}$ is
given by:
\begin{eqnarray}
\beta_{rp} = \frac{K Q_{pr}}{\rho_d a_d},
 \label{eq:eq3}
\end{eqnarray}
\noindent where $a_d$ and $\rho_d$ are the particle radius and the mass
density in the {\it MKS} system, and $K$ = 5.7 $\times$ 
10$^{-4}$ kg m$^{-2}$ is a constant. $Q_{pr}$ is a radiation pressure coefficient
the value of which depends on grain size, shape, structure and the optical constants of the grain material
\citep{Burns1979}. 

We applied a three-dimensional analysis to match the observed
images, following the model in \citet{Ishiguro2007},
\citet{Hanayama2012}, and \citet{Ishiguro2013}.
We adopted a power-law function for the
 terminal speed of ejected dust particles: 


\begin{equation}
V_{ej} = V_0
\left(\frac{\beta_{rp}}{\beta_{rp,0}}\right)^{u_1} v   ,
 \label{eq:eq4}
\end{equation}

\noindent
where $V_0$ is the reference ejection speed of particles having
$\beta_{rp,0}=1$ and $u_1$ is the power index of the ejection speed.
In a real comet the ejection speed will depend not only on
 $\beta_{rp}$ but also on the location of the dust source
on the nucleus, on the shape and porosity of the dust particles and perhaps on the ejection time within the outburst. 
The random variable $v$ 
in Eq. (\ref{eq:eq4})  reflects these uncertain factors.
It follows the Gaussian probability density function,
$P(v)$, 

\begin{equation}
P(v) = \frac{1}{\sqrt{2\pi} \sigma_v}\exp \left[-
	 \frac{(v-1)^2}{2\sigma_v^2}\right] , 
 \label{eq:eq5}
\end{equation}

\noindent
where $\sigma_v$ is the standard deviation of $v$.
In our computations, we limited the range $v-1<2\sigma_v$
in order to avoid very fast particles. In addition, we set the minimum ejection
speed to zero.

The number of dust particles at a given size  is written:

\begin{equation}
N(a_d;t)~da_d = N_0 \left(\frac{a_d}{a_{0}}\right)^{-q}~da_d, 
 \label{eq:eq6}
\end{equation}

\noindent
 in the size range of $a_{min}$ $\le$ $a_d$ $\le$ $a_{max}$,
 where $a_{min}$ and $a_{max}$ are minimum and maximum particle size
 given by $a_{min}=0.57/\rho_d \beta_{max}$  and $a_{max}=0.57/\rho_d\beta_{min}$,
 respectively, and $q$ is the power-index of the differential size distribution.


We imposed several constraints on the model.
First, we considered that all dust particles were released 
impulsively on UT 2010 November 2, neglecting the possibility of weaker dust ejection before and after this date. This assumption is supported by our
synchrone analysis and by the coma photometry as described above.
Secondly, we supposed that ejected dust particles are compact in
shape and can be represented by  $Q_{pr}$ = 1. This is a
reasonable approximation for optically large (2$\pi a_d/\lambda\ga$1, where $\lambda
\sim$0.64\micron~ is the wavelength) particles but is not strictly valid for optically
small particles ($a_d\la$0.2--0.3 \micron) \citep[see,
e.g.,][]{Ishiguro2007}. The dust mass density was assumed to be $\rho_d$=1000
kg m$^{-3}$. We also assumed that the dust particles were ejected symmetrically
with respect to the Sun--comet axis in a cone-shaped jet with a
half-opening angle $w$, implying that the explosion occurred around the
subsolar point of the nucleus.  Finally, we assumed that, for particles of all sizes, the geometric albedo
is 0.04 and the phase coefficient is $\beta$ = 0.035 mag deg$^{-1}$. 

We examined several key properties with which to constraint our dust model
from the observed images. We noticed that the envelope has a more open shape in the anti-solar direction 
meaning that the width of the envelope was enlarged by increasing ejection speeds even as the envelope was
stretched by the solar radiation pressure. Because smaller particles are more susceptible to  radiation
pressure, the envelope morphology suggests that small particles were ejected with higher speeds (see
Figure 3 and 4 (a)). From Eq. (\ref{eq:eq4}), we can derive the power index of the ejection speed for the
particles in the envelope, $u_1$=$\log(w_1/w_2)$/$\log(\beta_1/\beta_2)$, where $w_1$ and $w_2$ are the
apparent width of the envelope (proportional to the ejection speed projected on the celestial plane).
We examined the width and the corresponding $\beta_{rp}$ values from the image taken on 2010 November 9,
finding that $u_1$=0.30$\pm$0.05  best fits the observed broadening of the envelope.

Separately, we found that the envelope did not extend more than $\sim$4.5\arcmin~in our data. Particles with $\beta_{rp}>$2.5 should have spread to the edge of the field of view in the time since ejection, while particles with $\beta_{rp}<$1 would not match the observed extent.
Through a test simulation for hemispherical ejection model (e.g. Reach et al. (2010) section 6.1),
we obtained $\beta_{rp}\sim$1.5.
In the image on February, there is no obvious gap between the dust tail and the inner coma.
From the evidence, we put the upper limits of $\beta_{min}$ $\sim$1$\times$10$^{-3}$.

\begin{figure}
 \epsscale{1.0}
   \plotone{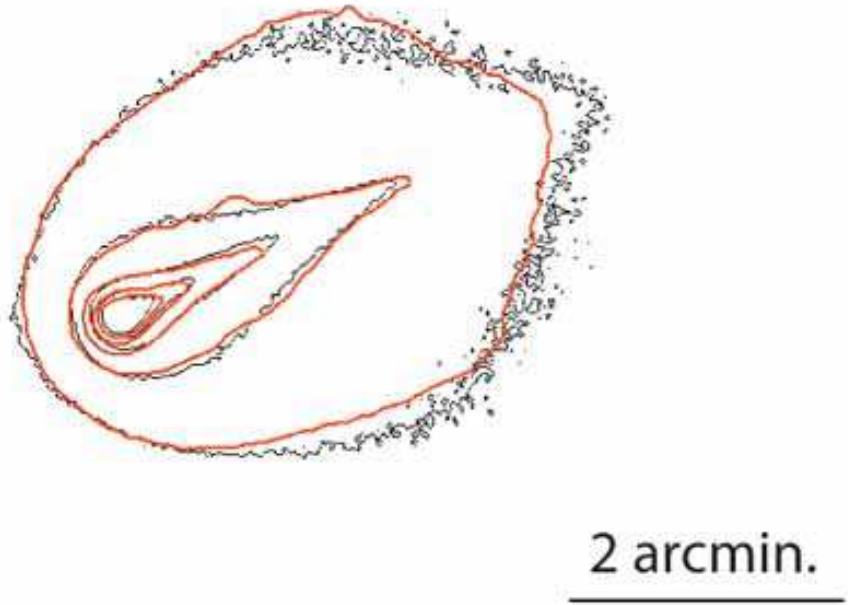}
  \caption{Comparison between the observed $R_\mathrm{C}$-band image (thin lines)
  and one of our best-fit models (thick lines) on  UT 2010 November 9. In this model, we used
  $u_1$=0.3, $q$=4.0, $\beta_\mathrm{min}$=0.5, $\beta_\mathrm{max}$=1.8, $V_0$=420 m s$^{-1}$,
  $\sigma_v$=0.05 and $w$=30\arcdeg for the envelope, 
    $u_1$=0.52, $q$=3.8, $\beta_\mathrm{min}$=1$\times$10$^{-3}$, $\beta_\mathrm{max}$=1.5,
  $V_0$=320 m s$^{-1}$,  $\sigma_v$=0.60 and $w$=35\arcdeg for the tail and coma.
  The contour levels are chosen arbitrary but the intervals are constant in linear scale.
  The field of view of these images is 9\arcmin$\times$6\arcmin.
 } 
 \label{fig:model_image}     
\end{figure}

Model images were produced in a Monte Carlo simulation by solving Kepler's
equation including solar gravity and radiation pressure. We derived the
above parameters to fit the surface brightness of the dust cloud on UT 2010
November 9, where prominent features (the envelope, tail and coma)
were detected. We created a number of simulation images using a wide
range of parameters as listed in Table \ref{tab:parameter}, and fitted the
image from the outer parts to the inner parts. A two-component
(i.e.~envelope and tail+coma) model worked well for the fitting. 
We selected 20 sampling points in the envelope and found the optimum
parameter sets first (envelope model). Then we subtracted the best-fit
envelope model from the observed intensity, and selected 25 sampling
points in the residual image, and derived the best-fit parameters to fit
the tail and coma surface brightness (tail+coma model). The best-fit
parameters are shown in Table \ref{tab:parameter}. We tolerate 
intensity differences between the model and observation of  up to 10\%, and derived the
errors in the Table. Figure \ref{fig:model_image} shows the comparison
between the observation and model. We produced the model contour
through further tuning of the best-fit parameters within the error range.
The distinctive morphology of the dust cloud is successfully reproduced by this two component model.

The best-fit parameters suggest that the envelope consists of small
particles ($\beta_{rp}$=0.5--1.8 or $a_d$=0.3--1 \micron)  with ejection speeds
higher than in the coma and tail. The reference speed of particles in the
envelope was $V_0$=420$\pm$30 m s$^{-1}$.  With the range of
$\beta_{pr}$, the ejection speed of the envelope particles turned out to
be 290--500 m s$^{-1}$, where we adopted $\sigma_v$=0 to derive the typical speed.
 On the other hand, the tail and coma consisted
of a wide range of dust particles  from sub-micron to
sub-millimeter ($\beta_{rp}$=1$\times$10$^{-3}$--1.5 or 0.4--570 \micron) in size. Their
ejection speeds are estimated to vary from 
7--390 m s$^{-1}$. 
The effective radius, $a_e$, of dust particles in the coma is given by
$a_e\approx\sqrt{0.4\times570}$ = 15 \micron. The ejection speed of
15 \micron-particle is 52$\pm$3 m s$^{-1}$ from Eq. (\ref{eq:eq4}),
which is fast enough to reach the projected radius of 15,000 km during
the time of our observation. This explains why the free
expansion model can characterize the observed magnitude profile
(Section \ref{sec:free_expansion}). 

We obtained the power index of  $\beta_{rp}$-dependence of the ejection
speed, $u_1$ = 0.30$\pm$0.05 in the envelope and 0.55$\pm$0.10 in the tail and coma.
Given the uncertainties, it is not clear that the difference between these estimates is formally
significant.  We note that the value $u_1\sim$ 0.5 is expected of dust particles
accelerated by  gas drag forces \citep{Whipple1951}.
The moderate slope for the envelope particles may suggest that 
small particles may be largely accelerated to reach the gas velocity.

We deduced the total mass of dust  and the total kinetic energy
by integrating with respect to particle size, as summarized in Table
\ref{tab:summary}. The total dust mass is $M_d$=5.1$\times$10$^8$kg.
With uncertainties in dust size ($a_{min}$ and $a_{max}$) and its power index ($q$) as well as the photometric error ($m_R$),
the derived mass is good to within a factor of four.
The dust mass corresponds to a  body 62-m in radius assuming 
mass density of $\rho_n$=500 kg m$^{-3}$. This  is $>$0.004 \% of
the  mass of a $r_n <$ 1850 m spherical body (the upper-limit of the nuclear
radius). The total kinetic energy  is $E_k$ = 5.0$\times$10$^{12}$ J, or 1.2
kiloton of TNT, with the bulk of the energy carried by the tail and coma particles.  Presumably, a comparable or larger energy was carried by gas in the initial explosion. The energy per unit mass is $E_k/M_d \sim$ 1$\times$10$^4$ J
kg$^{-1}$. The value is similar to that of 17P/Holmes
\citep{Li2011,Reach2010} and is about 10\% of the energy released by the crystallization of amorphous water ice
(9$\times$10$^4$ J kg$^{-1}$). 

The ejected mass could be
contained in a surface layer on the nucleus having thickness \citep[see,
e.g.,][]{Li2011}, 

\begin{equation}
l = \frac{M_d}{4 \pi r_n^2 f \rho_n},
 \label{eq:eq7}
\end{equation}

\noindent
where $f$ is the fraction of the surface area of the nucleus that is
ejected. We obtained $w$ =30--35\arcdeg~ to an accuracy of
$\sim$10\degr~from our model simulations, which
suggests that the active area exists within $w \lesssim$30\arcdeg~ from the
sub-solar point. The area of the inferred active region is 2.9$\times$10$^6$
m$^2$, corresponding to $f$=0.07. Substituting these values gives $l >$0.35 m.  The ejected mass could be contained within a circular patch of the nucleus surface roughly 1 km in radius and 35 cm thick.

\subsection{Dynamical Evolution of the Nucleus}

Here we examine the orbital evolution of V1 to attempt to understand its recent history. Dynamical chaos imposes a fundamental limit to our ability to backwards-integrate the motion of any comet; a small error in the initial conditions will grow
exponentially on the Lyapunov time. There is additional
uncertainty from the (generally poorly known) non-gravitational acceleration, which is induced in comets by
recoil forces from the sublimation of ice.   The non-gravitational parameters of V1 are not known.  In the case of V1, there is in addition a relatively large
uncertainty in the orbital elements because these were necessarily determined from observations taken over a  short  interval (only 80 days).

To investigate the past orbit, we consider many `clones', whose initial orbits follow a Gaussian
distribution with the average values and the standard deviations
provided by the NASA/JPL HORIZONS site (Table \ref{tab:orbital_elements}). Then the
clone orbits are calculated and examined statistically. We
generated 1,000 clones of V1  using the N-body
integration package, Mercury \citep{Chambers1999}, and calculated the orbital evolution over
the past 10,000 years.  We set the non-gravitational force equal to zero.

\begin{figure}
 \epsscale{0.8}
   \plotone{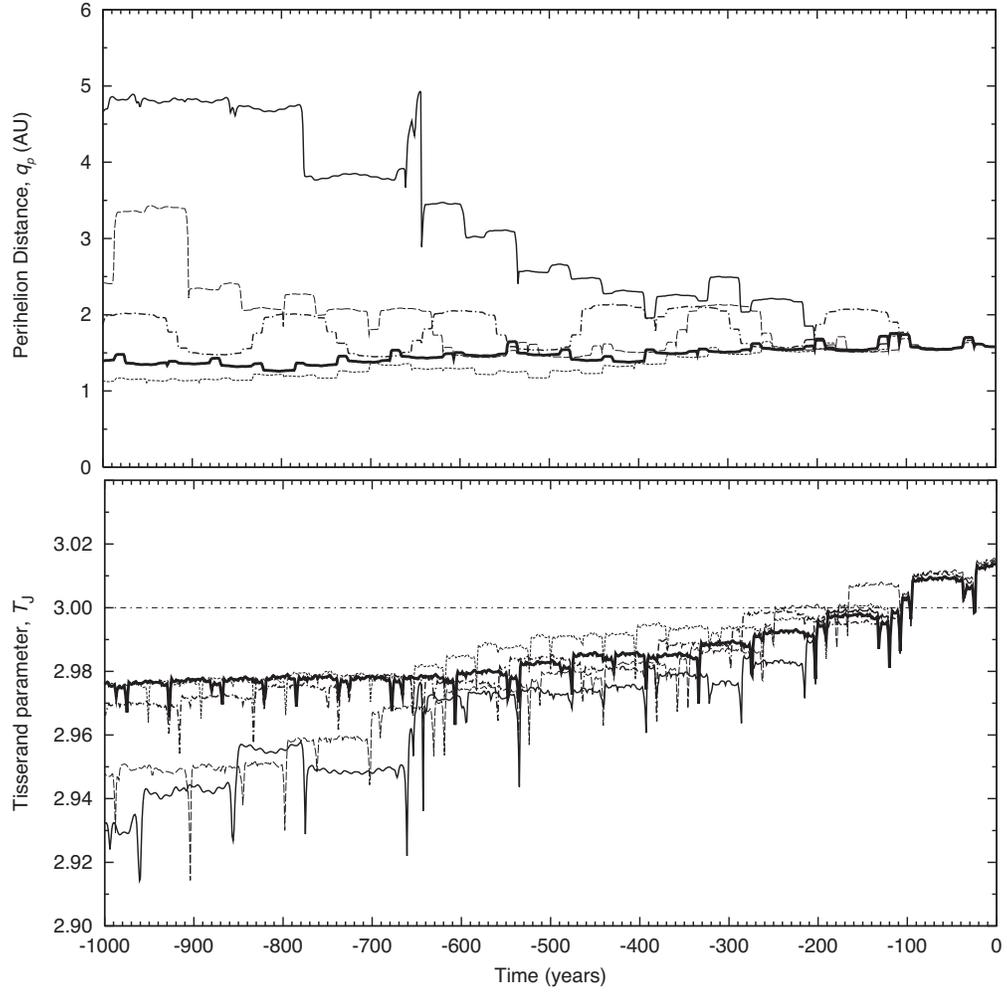}
  \caption{Examples of orbital evolution of V1 clone.}
 \label{fig:orb_evo}     
\end{figure}

Figure \ref{fig:orb_evo} shows  the orbital
evolution of five sample
clones. They follow almost identical orbits for about 100 years before present epoch, with perihelion fixed near 1.6 AU. Their Tisserand parameters drop below 3
and become Jupiter-family comets within 100--200 years. Thus, V1 is likely to be a Jupiter-family comet which  originated in the
Kuiper-belt region. Comets generally become active within $\approx$2.5 AU owing to
sublimation. We examined the fraction of V1 clones which existed
within 2.5 AU as visible comets. We found that all the V1 clones had perihelion $<$2.5 AU
over the last 100 years, dropping to 74\% over 1,000 years  and 19\%  in 10,000 years. On this basis,
it is clear that V1 is unlikely to be  a new comet 
making its first appearance at small heliocentric distances.  Therefore, the non-detection of V1 before 2010 is either a result of sky-survey incompleteness (unlikely, given the brightness of the comet) or a reflection of much reduced activity in previous orbits.  We conjecture
that, until the outburst on 2010 November 2, activity on the nucleus was largely stifled  by a dust mantle, leading to low brightness and the non-detection of V1.

\subsection{COMPARISON WITH  OTHER COMETS}

\begin{figure}
 \epsscale{1.0}
   \plotone{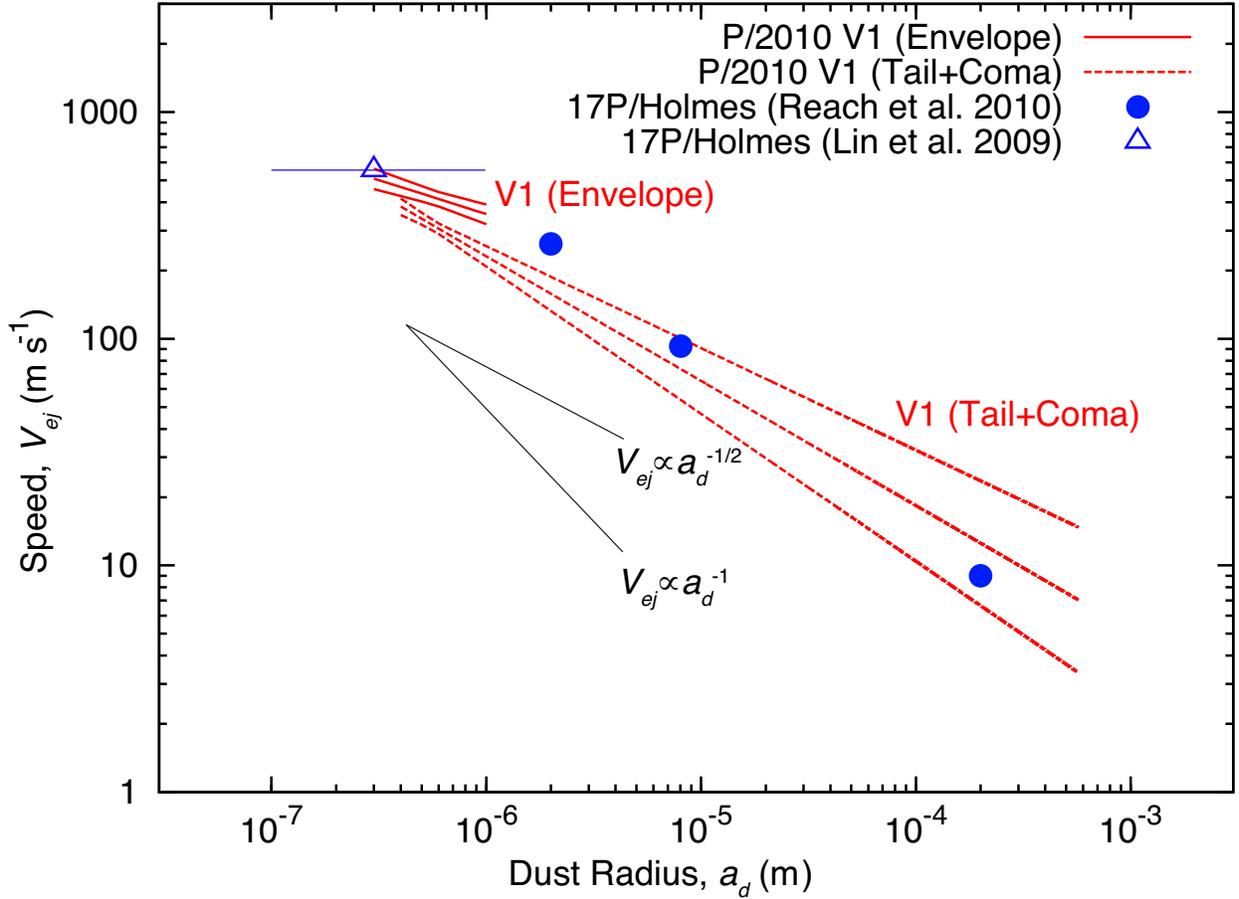}
  \caption{Comparison of speed between 17P/Holmes event in 2007 and V1
 event in 2010. Three lines for V1 denote maximum, nominal, and minimum
 speed based on our model simulation (see $V_0$ and $u_1$ in Table 4. $\sigma_v$ is not considered
 in this graph). Three filled circles are obtained from
 \citet{Reach2010}. Open triangle is the projected speed of dust envelopes
 observed soon after the outburst \citep{Montalto2008,Lin2009}, where we
 assumed the particles size of sub-micron (i.e.~0.1--1 \micron).}
 \label{fig:speed}     
\end{figure}

Like V1, 17P/Holmes was discovered (in 1892) because of a dramatic outburst.  Another  outburst, in 2007, was well observed, revealing a
spherical envelope, a detached blob, and a central  coma \citep[see,
e.g.,][]{Watanabe2009,Reach2010}. Total ejecta mass was estimated to be
(1$\sim$610)$\times$10$^{10}$ kg 
\citep{Altenhoff2009,Reach2010,Ishiguro2010,Li2011,Boissier2012,Ishiguro2013}.
The expansion speed on the plane of the sky of the dust envelope
particles was 554$\pm$5 m s$^{-1}$ \citep{Lin2009,Montalto2008}.
Several other comets are known to have undergone huge photometric
outbursts  accompanied by circular envelopes. For example, 41P/Tuttle-Giacobini-Kresak
experienced an outburst at 1.15 AU, and, before fading
underwent second outburst at 1.25 AU from the Sun. It possessed an
envelope (probably consisting of dust and gas \citep{Sekanina2008a}) expanding at 300 --700 m s$^{-1}$ \citep{Kresak1974}. 1P/Halley experienced a massive
explosion in 1836 at 1.44 AU from the Sun. Similarly, 1P/Halley was
enclosed by a circular envelope consisting of dust particles traveling at a speed
of 575$\pm$9 m s$^{-1}$  \citep{Sekanina2008b}. Only 
17P/Holmes  and V1  were observed with modern
astronomical instruments (i.e.~CCD) and the others were observed by
photographic plates or naked eyes. We summarize the physical quantities
of the outburst events at 17P/Holmes and V1 in Table
\ref{tab:comparison}. 
Although the magnitudes and heliocentric distances are 
different, the maximum speeds are similar to one another.
Figure \ref{fig:speed} shows the comparison between the 2010 V1 event (this
work) and the 2007 17P/Holmes event \citep{Reach2010,Lin2009}. The dust size
was not specified in \citet{Lin2009} and \citet{Montalto2008}, but we
regard  
it as sub-micron particles (i.e.~0.3$^{+0.7}_{-0.2}$\micron) because
only such small particles can be  accelerated to the highest
velocity and remain as sensitive scatterers in optical observations.
\citet{Reach2010} provided the speeds for three different 
populations (core, blob and shell). Although the total dust mass and the
kinetic energy of these two events are different, the size--speed
relationships are quite similar to one another.

Several possible mechanisms have been presented to explain 17P/Holmes
outburst; these include vaporization of pockets of 
more volatile ices such as CO$_\mathrm{2}$ and CO 
\citep{Schleicher2009,Kossacki2011}, the phase change of water from
amorphous to crystalline ice \citep{Sekanina2009}, thermal stress in the
nucleus, or the polymerization of hydrogen cyanide
\citep{Gronkowski2010}. A plausible trigger is the
crystallization of amorphous water ice \citep{Prialnik2004}.
From Table \ref{tab:comparison}, most of large-scale outbursts occurred
after their perihelion passages, suggesting that a time-lag from conducted heat might trigger these
outbursts. 


The heat diffusion equation can be solved to give the distance
over which heat can be transported by conduction, 
$\delta r = (\kappa P / \pi)^{1/2}$, where $\kappa$ is the thermal
diffusivity of the surface materials and $P$ is the period of time over
which conduction acts \citep{Li2011}. The applicable thermal diffusivity in comets is uncertain, depending on the unknown porosity of the material.  Insulating solids typically have $\kappa \sim$ 10$^{-6}$ m$^2$ s$^{-1}$ while  $\kappa = 10^{-7}$ to $10^{-8}$ m$^2$ s$^{-1}$ maybe more appropriate for comets in which porous structure reduces the contact area between grains \citep{Prialnik2004}. If, as seems likely from the clone experiments, V1 has spent $\gtrsim$100 yr inside 2.5 AU,  conducted heat
would reach a depth $\delta r \gtrsim$ 3 to 10 m beneath the initial surface. Since $\delta r \gtrsim l$ (Equation \ref{eq:eq7}), it is quite plausible, although far from proved,  that the outburst was
triggered by the action of conducted heat through the crystallization of
buried amorphous ice.

\clearpage

\section{SUMMARY}

From our research on V1, we find the following.

\begin{enumerate}
\item{

     Several  observations show that V1 underwent an explosive ejection in late 2010.  The changing position angle of the dust tail
     is closely matched by synchrone trajectories for ejection dates between UT 2010 October 31 and November 3. The  near-nucleus coma faded  steadily at $\sim$0.06 mag day$^{-1}$, distinct from any steady-state behavior.   The non-discovery of this nearby, bright comet before 2010 is naturally explained by the outburst interpretation.
}

%

\item{

The  V1 dust cloud had two distinct components.
The envelope consisted of small grains (radii 0.3--1 \micron)
     expanding at a maximum speed of 500$\pm$40 m s$^{-1}$. The tail
     and coma were composed of a wider range of dust particles (radii 0.4--570
     \micron) and  expansion speeds of 7--390 m s$^{-1}$.
}
\item{

The ejecta  mass in solids is 5.1$\times10^{8}$ kg and the kinetic
     energy is  5.0$\times$10$^{12}$ J.  Although the mass and energy are
     orders of magnitude smaller than in 17P/Holmes, the
     energy per unit mass ($\sim$1$\times$10$^4$ J kg$^{-1}$) is similar.
     
}

\item The sudden ejection and the derived energy per unit mass of the ejecta are consistent with runaway crystallization of buried amorphous ice as the source of energy to drive the outburst.

\end{enumerate}

\vspace{1cm}
{\bf Acknowledgments}\\
We would like to express our gratitude for vigorous activity and prompt
reactions by amateur astronomers' network. MI was supported by a National
Research Foundation of Korea (NRF) grant funded by the Korean government
(MEST) (No. 2012R1A4A1028713). 
This research was partially supported by the Ministry of Education, Culture,
Sports, Science and Technology of Japan (MEXT), Grant-in-Aid No. 14GS0211,
and 19047003.
Data collected at Kiso Observatory was
obtained from the SMOKA, which is operated by the Astronomy Data Center,
National Astronomical Observatory of Japan.


\clearpage

\clearpage

\clearpage

\clearpage

\clearpage

\clearpage

\clearpage

\clearpage

\clearpage

\clearpage

\clearpage

\begin{table}
\footnotesize
  \caption{Observation Log.}  
 \label{tab:cirsumstance}
  \begin{center}
    \begin{tabular}{cccccccc}
     \hline
DATE & UT & Observatory (Instrument) & Filter (Exptime\tablenotemark{a})
     & Seeing\tablenotemark{b} & $r_h$\tablenotemark{c} & 
     $\Delta$\tablenotemark{d} & $\alpha$\tablenotemark{e} \\   
 \hline

2010--11--09 & 20:32--21:02 & IAO (MITSuME) & $g'$ (26), $R_\mathrm{C}$
		     (27), $I_\mathrm{C}$ (27) & 3.4 & 1.60 & 2.32 & 20.3 \\
2010--11--20 & 20:25--21:12 & IAO (MITSuME) & $g'$ (26), $R_\mathrm{C}$
		     (38), $I_\mathrm{C}$ (38) & 2.5 & 1.62 & 2.29 & 21.7 \\
2010--11--26 & 20:24--21:14 & IAO (MITSuME) & $g'$ (28), $R_\mathrm{C}$
		     (45), $I_\mathrm{C}$ (45) & 4.2 & 1.63 & 2.27 & 22.5 \\
2010--12--09 & 20:13--21:23 & IAO (MITSuME) & $g'$ (66), $R_\mathrm{C}$
		     (66), $I_\mathrm{C}$ (66) & 3.3 & 1.67 & 2.24 & 24.1 \\
2010--12--12 & 20:25--21:12 & IAO (MITSuME) & $g'$ (42), $R_\mathrm{C}$
		     (42), $I_\mathrm{C}$ (42) & 2.4 & 1.68 & 2.23 & 24.4 \\
2010--12--19 & 20:37--21:28 & IAO (MITSuME) & $g'$ (48), $R_\mathrm{C}$
		     (48), $I_\mathrm{C}$ (48) & 2.5 & 1.70 & 2.20 & 25.2 \\
2011--01--30 & 15:48--15:55 & KECK-I (LRIS) & $B$ (4.6), $R_\mathrm{C}$
		     (3.7) & 1.0 & 1.87 & 2.03 & 28.9 \\
2011--02--04 & 19:41--21:25 & IAO (MITSuME) & $g'$ (80), $R_\mathrm{C}$
		     (80), $I_\mathrm{C}$ (80) & 4.8 & 1.90 & 2.00 & 29.1 \\
2011--03--29 & 21:44--23:28 & HCT (HFOSC)  & $R_\mathrm{C}$ (63)
		                               & 3.0 & 2.17 & 1.68 & 26.3 \\
\hline
    \end{tabular}
   \tablenotetext{a}{Total effective exposure time in minutes.}
   \tablenotetext{b}{FWHM seeing in arcsec.}
   \tablenotetext{c}{Median heliocentric  distance in AU.}
   \tablenotetext{d}{Median geocentric  distance in AU.}
   \tablenotetext{e}{Median Solar phase angle (Sun--V1--Observer angle)in degree.}
  \end{center}
\end{table}
\clearpage

\begin{table}
  \caption{$R_\mathrm{C}$-band Photometric Results.}
  \begin{center}
    \begin{tabular}{lrrrrrrr}
\hline
Median Time (UT) & $m_R$ [error$^\dag$] & $m_R(1,1,0)$ \\
\hline
2010-11-09.87 & 13.45 [0.20] &  9.90 \\
2010-11-20.87 & 14.62 [0.20] & 11.02 \\
2010-11-26.87 & 15.38 [0.25] & 11.75 \\
2010-12-09.87 & 16.00 [0.25] & 12.29 \\
2010-12-12.87 & 16.56 [0.30] & 12.83 \\
2010-12-19.88 & 16.50 [0.20] & 12.75 \\
2011-01-30.66 & 18.93 [0.07] & 15.02 \\
2011-02-04.86 & 19.07 [0.60] & 15.15 \\
2011-03-29.94 & $>$20.00 & $>$16.27 \\
\hline
    \end{tabular}
  \end{center}
 \label{tab:photometry}
\tablecomments{$^\dag$ magnitude error 1 $\sigma$}
\end{table}
\clearpage

\begin{table}
  \caption{Observational circumstance and $R_\mathrm{C}$-band
 Photometric Results of 17P/Holmes.} 
  \begin{center}
    \begin{tabular}{lrrrrrrr}
\hline
Median Time & $r_h$ & $\Delta$ & $\alpha$ & $m_R$ [error$^\dag$] &
     $m_R(1,1,0)$ \\ 
\hline
2007-10-27.66 & 2.45 & 1.63 & 16.10 & 7.50 [0.31]  & 3.93 \\
2007-10-30.71 & 2.46 & 1.62 & 15.30 & 8.56 [0.31]  & 5.02 \\
2007-11-03.69 & 2.48 & 1.62 & 14.40 & 10.14 [0.31]  & 6.62 \\
2007-11-07.63 & 2.49 & 1.62 & 13.50 & 10.73 [0.31]  & 7.23 \\
2007-11-11.61 & 2.51 & 1.62 & 12.60 & 11.42 [0.55]  & 7.93 \\
2007-11-13.62 & 2.52 & 1.63 & 12.30 & 11.97 [0.31]  & 8.47 \\
2007-11-18.53 & 2.54 & 1.64 & 11.60 & 12.62 [0.32]  & 9.11 \\
2007-11-22.43 & 2.55 & 1.65 & 11.30 & 13.00 [0.68]  & 9.48 \\
2007-12-01.55 & 2.59 & 1.69 & 11.40 & 13.39 [0.31]  & 9.78 \\
2007-12-13.58 & 2.64 & 1.78 & 12.90 & 14.24 [0.34]  & 10.42 \\
2007-12-16.51 & 2.65 & 1.81 & 13.40 & 13.93 [0.31]  & 10.06 \\
2008-02-07.57 & 2.88 & 2.54 & 19.70 & 15.82 [0.32]  & 10.81 \\
2008-02-28.52 & 2.97 & 2.90 & 19.40 & 16.64 [0.58]  & 11.29 \\
\hline
    \end{tabular}
  \end{center}
 \label{tab:17Pphotometry}
\tablecomments{$^\dag$ magnitude error 1 $\sigma$}
\end{table}
\clearpage

\begin{table}
  \caption{Dust Model Parameters}
  \begin{center}
    \begin{tabular}{lllll}
\hline
Parameter   & Input values & Best-fit (Envelope) & Best-fit (Tail+Coma)
     & Unit\\
\hline
$u_1$ & 0.1--0.9 with 0.1 interval & 0.3 (fixed) & 0.55$\pm$0.1 & --\\
$q$ & 3.0--4.5 with 0.1 interval & 4.0$\pm$0.5 & 3.8$\pm$0.1 & --\\
$\beta_\mathrm{max}$ & 1.2, 1.5, 1.8  &
	     1.8 & 1.5  & --\\
$\beta_\mathrm{min}$ & 0.5, 1$\times$10$^{-1}$, 1$\times$10$^{-2}$, 1$\times$10$^{-3}$
	 & 0.5 & 1$\times$10$^{-3}$ (fixed) & --\\
$V_0$ & 150--600 with 30 interval & 420$\pm$30 & 315$\pm$15 & m s$^{-1}$\\
$\sigma_v$ & 0--0.5 with 0.1 interval & 0.1$\pm$0.05 & 0.5$\pm$0.1 & --\\
$\omega$ & 5--60 with 5 interval & 30$\pm$5 & 35$\pm$10 & degree\\
\hline
    \end{tabular}
  \end{center}
 \label{tab:parameter}
\end{table}
\clearpage

\begin{table}
  \caption{Derived Physical Characteristics}
  \begin{center}
    \begin{tabular}{lllll}
\hline
Quantity        & Envelope  & Tail+Coma & Total & Unit\\
\hline
Speed$^\dag$    & 420$\pm$30& 315$\pm$15   & --   & m s$^{-1}$\\
Particle radius & 0.3--1    & 0.4--570     & --   & 10$^{-6}$ m\\
Cross Section   & 3.2$\pm$0.3 & 7.2$\pm$0.7 & 10.4$\pm$1.0  & 10$^{10}$ m$^2$\\
Mass            & 0.24      & 4.84         & 5.1  & 10$^{8}$ kg\\
Kinetic Energy  & 2.2       & 2.8          & 5.0  & 10$^{12}$ J\\
\hline
    \end{tabular}
  \end{center}
\tablecomments{$^\dag$ The speed of grains having $\beta$=1 (radius 0.57
 \micron~for density $\rho$ = 1000 kg m$^{-3}$). }
 \label{tab:summary}
\end{table}
\clearpage

\begin{table}
  \caption{Orbital elements (Epoch 2455518.5, UT 2010-Nov-18.0)}
  \begin{center}
    \begin{tabular}{lrrrr}
\hline
Element	& Value	& Uncertainty (1$\sigma$) & Unit \\
\hline
eccentricity, $e$	& 0.48803	& 0.00022 & \\	 
semi-major axis, $a$	& 3.0832	& 0.0016  &	AU \\
perihelion distance, $q_p$ &1.57854	& 0.00013 &	AU \\
inclination, $i$	& 9.37832	& 0.00018 &	degree \\
longitude of the ascending node, $\Omega$& 3.8155 & 0.0013 &degree \\
argument of perihelion, $\omega$ & 152.396	& 0.014   &degree \\
time of perihelion passage, $T_p$& 2455482.783  & 0.022   &	JED \\
\hline
    \end{tabular}
  \end{center}
 \label{tab:orbital_elements}
\end{table}
\clearpage

\begin{table}
\footnotesize
  \caption{Comparison Between P/2010 V1 and 17P/Holmes Outbursts}
  \begin{center}
     \begin{tabular}{lrrrr}
\hline
Quantity     & P/2010 V1 & 17P/Holmes & References for 17P/Holmes\\
\hline
$a$\tablenotemark{1} & 3.083     & 3.621 \\
$e$\tablenotemark{2} & 0.488     & 0.432 \\
$i$\tablenotemark{3} & 9.378     & 19.090 \\
$q_p$\tablenotemark{4} & 1.579     & 2.057  \\
$r_h$\tablenotemark{5} & 1.59      & 2.44 \\
$r_N$\tablenotemark{6} & $<$1.85   & 1.71 & \citet{Lamy2004}\\
$\Delta t_p$\tablenotemark{7} & +20       & +172  & \citet{Hsieh2010}\\
$m_R(1,1,0)$\tablenotemark{8} & 5.97$\pm$0.14 & -1.12$\pm$0.30 & This work\tablenotemark{16}\\
$A$\tablenotemark{9}  & (1.0$\pm$0.2)$\times$10$^{11}$ &
	      (7.1$\pm$2.2)$\times$10$^{13}$ & This work\tablenotemark{16}\\
$t_{rise}\tablenotemark{10}$ & $\approx$1 & 1.2$\pm$0.3 & \citet{Li2011}\\
$t_{fade}\tablenotemark{11}$            & 70 & 50 & \citet{Stevenson2012}\\
$M_d$\tablenotemark{12}   & 5.1$\times$10$^{8}$ &
	      (1$\sim$610)$\times$10$^{10}$ & \citet{Li2011,Ishiguro2013}\\
$V_{max}$\tablenotemark{13}   & 500$\pm$40 & 554$\pm$5 & \citet{Lin2009}\\
$E_k$\tablenotemark{14}     & 5.0$\times$10$^{12}$ &
	      (1.2$\sim$1400)$\times$10$^{14}$ & \citet{Li2011,Reach2010}\\
$E_k/M_d$\tablenotemark{15} & 1$\times$10$^{4}$  & 1.2$\times$10$^4$
	      & \citet{Reach2010}\\
\hline
     \end{tabular}
   \tablenotetext{1}{Semi-major axis in AU.}
   \tablenotetext{2}{Eccentricity.}
   \tablenotetext{3}{Inclination in degree.}
   \tablenotetext{4}{Perihelion distance in AU.}
   \tablenotetext{5}{Heliocentric distance at the time of outburst in AU.}
   \tablenotetext{6}{Radius of nucleus in km.}
   \tablenotetext{7}{Onset time after perihelion passage in days.}
   \tablenotetext{8}{Absolute $R_\mathrm{C}$-band magnitude.}
   \tablenotetext{9}{Total cross section of dust cloud in m$^2$.}
   \tablenotetext{10}{Rise time in days.}
   \tablenotetext{11}{Fade time when the magnitude decreased by 4 mag in days.}
   \tablenotetext{12}{Ejecta mass in kg.}
   \tablenotetext{13}{Maximum speed of ejecta in m s$^{-1}$.}
   \tablenotetext{14}{Kinetic energy in J.}
   \tablenotetext{15}{Kinetic energy per unit mass in J kg$^{-1}$.}
   \tablenotetext{16}{These were obtained by ourselves using images taken at Kiso observatory.}
  \end{center}
 \label{tab:comparison}
\end{table}
\clearpage


\begin{thebibliography}{}
\bibitem[Altenhoff et al.(2009)]{Altenhoff2009}
Altenhoff, W. J., Kreysa, E., Menten, K. M., Sievers, A., Thum, C., \&
Weiss, A.\ 2009, \aap, 495, 975 

\bibitem[Boissier et al.(2012)]{Boissier2012} Boissier, J.,
 et al.\ 2012, \aap, 542, A73  

\bibitem[Brown et al.(1996)]{Brown1996} Brown, M.~E., Bouchez, 
A.~H., Spinrad, A.~H., \& Johns-Krull, C.~M.\ 1996, \aj, 112, 1197 

\bibitem[Burns et al.(1979)]{Burns1979} Burns, J. A., Lamy, 
P. L., \& Soter, S.\ 1979, Icarus, 40, 1 

\bibitem[Capria et al.(2010)]{Capria2010} Capria, M.~T.,R
Cremonese, G., \& de Sanctis, M.~C.\ 2010, \aap, 522, A82  

\bibitem[Chambers(1999)]{Chambers1999} Chambers, J.~E.\ 1999, 
\mnras, 304, 793 

\bibitem[Combi \& Delsemme(1980)]{Combi1980} Combi, M.~R., \& Delsemme,
		A.~H.\ 1980, \apj, 237, 633  

\bibitem[Gronkowski \& Sacharczuk(2010)]{Gronkowski2010}
Gronkowski, P., \& Sacharczuk, Z.\ 2010, \mnras, 408, 1207  

\bibitem[Hanayama et al.(2012)]{Hanayama2012} Hanayama, H., 
 et al.\ 2012, \pasj, 64, 134 

\bibitem[Hsieh et al.(2010)]{Hsieh2010} Hsieh, H.~H., 
Fitzsimmons, A., Joshi, Y., Christian, D., 
\& Pollacco, D.~L.\ 2010, \mnras, 407, 1784 

\bibitem[Holmberg et al.(2006)]{Holmberg2006} Holmberg, J., Flynn, 
C., \& Portinari, L.\ 2006, \mnras, 367, 449 

\bibitem[Jewitt et al.(2010)]{Jewitt2010} Jewitt, D., Weaver, H., 
Agarwal, J., Mutchler, M., \& Drahus, M.\ 2010, \nat, 467, 817 

\bibitem[Ishiguro et al.(2007)]{Ishiguro2007}
Ishiguro, M.,  Sarugaku, Y., Ueno, M., Miura, N., Usui, F., Chun, M.-Y., 
\& Kwon, S. M.\ 2007, Icarus, 189, 169 

\bibitem[Ishiguro(2008)]{Ishiguro2008}
Ishiguro, M.\ 2008, Icarus, 193, 96 

\bibitem[Ishiguro et al.(2010)]{Ishiguro2010} Ishiguro, M., 
 et al.\ 2010, \apj, 714, 1324 

\bibitem[Ishiguro et al.(2013)]{Ishiguro2013} Ishiguro, M., Kim, Y., 
Kim, J., et al.\ 2013, \apj, 778, 19

\bibitem[Kim et al.(2012)]{Kim2012} Kim, J., Ishiguro, M., 
Hanayama, H., et al.\ 2012, \apjl, 746, L11 

\bibitem[Kossacki \& Szutowicz(2011)]{Kossacki2011} Kossacki, K. J., \&
Szutowicz, S.\ 2011, \icarus, 212, 847  

\bibitem[Kresak(1974)]{Kresak1974} Kresak, L.\ 1974, Bulletin of 
the Astronomical Institutes of Czechoslovakia, 25, 293 

\bibitem[Lamy et al.(2004)]{Lamy2004}
Lamy, P. L., Toth, I., Fernandez, Y. R., \& Weaver, H. A.\ 2004, Comets
II, 223  

\bibitem[Landolt(1992)]{Landolt1992} Landolt, A. U.  1992, \aj, 104, 1, 340

\bibitem[Landolt(2009)]{Landolt2009} Landolt, A. U.\ 2009, \aj, 
137, 4186 

\bibitem[Levison \& Duncan(1997)]{Levison1997} Levison, H. F.,
                \& Duncan, M. J.\ 1997, \icarus, 127, 13  

\bibitem[Li et al.(2011)]{Li2011} Li, J., Jewitt, D., Clover, 
J. M., \& Jackson, B. V.\ 2011, \apj, 728, 31 

\bibitem[Lin et al.(2009)]{Lin2009} Lin, Z.-Y., Lin, C.-S., Ip, 
W.-H., \& Lara, L.~M.\ 2009, \aj, 138, 625 

\bibitem[Mink(1997)]{Mink1997} Mink, D.~J.\ 1997, Astronomical 
Data Analysis Software and Systems VI, 125, 249 

\bibitem[Monet et al.(2003)]{Monet2003} Monet, D. G., et al.\ 
2003, \aj, 125, 984 

\bibitem[Montalto et al.(2008)]{Montalto2008} Montalto, M., Riffeser,
A., Hopp, U., Wilke, S., \& Carraro, G.\ 2008, \aap, 479, L45 

\bibitem[Nakano \& Ikeya(2010a)]{Nakano2010a} Nakano, S., \&
Ikeya, K.\ 2010, \iaucirc, 9175, 1  

\bibitem[Nakano \& Ikeya(2010b)]{Nakano2010b} Nakano, S., \& Ikeya, K.\
2010, \iaucirc, 9183, 3 

\bibitem[Oke et al.(1995)]{Oke1995} Oke, J.~B., Cohen, J.~G., 
Carr, M., et al.\ 1995, \pasp, 107, 375 

\bibitem[Prialnik et al.(2004)]{Prialnik2004} Prialnik, D., 
Benkhoff, J., \& Podolak, M.\ 2004, Comets II, 359 

\bibitem[Reach et al.(2010)]{Reach2010} Reach, W. T., Vaubaillon, 
J., Lisse, C. M., Holloway, M., \& Rho, J.\ 2010, \icarus, 208, 276 

\bibitem[Schleicher(2009)]{Schleicher2009} Schleicher, D. G.\ 2009, \aj,
138, 1062 

\bibitem[Sekanina(1982)]{Sekanina1982} Sekanina, Z.\ 1982, IAU 
Colloq.~61: Comet Discoveries, Statistics, and Observational Selection, 251 

\bibitem[Sekanina \& Larson(1984)]{Sekanina1984} Sekanina, Z., \&
		Larson, S.~M.\ 1984, \aj, 89, 1408  

\bibitem[Sekanina(2008a)]{Sekanina2008a} Sekanina, Z.\ 2008, 
International Comet Quarterly, 30, 3 

\bibitem[Sekanina(2008b)]{Sekanina2008b} Sekanina, Z.\ 2008, 
Int. Comet Q. 30, 63 

\bibitem[Sekanina(2009)]{Sekanina2009}
Sekanina, Z.\ 2009, Int. Comet Q., 31, 5 

\bibitem[Stevenson \& Jewitt(2012)]{Stevenson2012} Stevenson, R., \&
Jewitt, D.\ 2012, \aj, 144, 138  

\bibitem[Watanabe et al.(2009)]{Watanabe2009} Watanabe, J.-I., et al.
2009, \pasj, 61, 679  


\bibitem[Whipple(1951)]{Whipple1951} Whipple, F.~L.\ 1951, \apj, 
113, 464 

\bibitem[Williams(2010)]{Williams2010} Williams, G.~V.\ 2010, 
\iaucirc, 9189, 2 

\bibitem[Yang et al.(2009)]{Yang2009} Yang, B., Jewitt, D., 
\& Bus, S. J.\ 2009, \aj, 137, 4538 

\end{thebibliography}
\end{document}